\documentclass[aps,prl,showpacs,preprint,superscriptaddress]{revtex4}
\usepackage{amsmath}
\usepackage{graphicx}

\begin{document}

\title{Quantum plasma effects in the classical regime}

\author{G. Brodin}
\affiliation{Department of Physics, Ume{\aa} University, SE--901 87 Ume{\aa},
Sweden}

\author{M. Marklund}
\affiliation{Department of Physics, Ume{\aa} University, SE--901 87 Ume{\aa},
Sweden}

\author{G. Manfredi}
\affiliation{Institut de Physique et Chimie des Mat\'eriaux
de Strasbourg, UMR 7504 ULP-CNRS, 23 Rue du Loess, BP 43, F--67034
Strasbourg Cedex 2, France}

\received{\today}

\begin{abstract}
For quantum effects to be significant in plasmas it is often
assumed that the temperature over density ratio must be small. In
this paper we challenge this assumption by considering the
contribution to the dynamics from the electron spin properties. As
a starting point we consider a multicomponent plasma model, where
electrons with spin up and spin down are regarded as different
fluids. By studying the propagation of Alfv\'{e}n wave solitons we
demonstrate that quantum effects can survive in a relatively
high-temperature plasma. The consequences of our results are
discussed.
\end{abstract}


\maketitle
Recently, several studies of quantum plasmas have appeared in the
literature \cite{manfredi,haas-etal1,haas,haas-etal2,garcia-etal,shukla,Shukla-Eliasson,marklund-brodin,MHD-spin,EP-spin,Ferro-spin,marklund-etal,brodin-etal,piazza,Lundin}, where 
the Bohm--de
Broglie potential and 
the Fermi pressure \cite{manfredi,haas-etal1,haas,haas-etal2,garcia-etal,shukla,Shukla-Eliasson},
spin properties \cite{marklund-brodin,MHD-spin,EP-spin,Ferro-spin,marklund-etal,Cowley,Kulsrud,Ferrofluid}
as well as certain quantum electrodynamical effects \cite
{brodin-etal,piazza,Lundin,RMP} are accounted for. The applications range from plasmonics \cite
{SPP} and quantum wells \cite{Quantum wells}, to ultracold plasmas \cite
{Ultracold} and astrophysics \cite{brodin-etal,Harding2006}. Quantum plasma
effects can also be seen in scattering experiments with solid density
targets \cite{Glenzer2007}. The usual regime
where quantum effects are important involves dense low-temperature
plasmas, where either the Fermi pressure is comparable to the
thermal pressure or the thermal de Broglie wavelength times the
plasma frequency is comparable to the thermal velocity. In recent
studies of spin effects in plasmas \cite
{marklund-brodin,MHD-spin,EP-spin,Ferro-spin,marklund-etal}, the condition for quantum
effects to be important has been found to be somewhat different from the case of
non-spin quantum plasmas \cite{manfredi,haas-etal1,haas,haas-etal2,garcia-etal,shukla,Shukla-Eliasson},
but also here a high temperature tend to make quantum effects
small.

In the present Letter we study a weakly collisional high
temperature plasma. In particular, we focus on the case where the
temperature over magnetic field ratio is sufficiently high to make
spins randomly oriented at thermodynamic equilibrium. Within the
one-fluid model, such a condition
tends to make the macroscopic spin effects negligible \cite
{marklund-brodin,MHD-spin,EP-spin,Ferro-spin}. However,
here we study a two-electron fluid model, where the different
electron populations are defined by their spin relative to the
magnetic field. Evaluating this model for the particular case of
Alfv\'{e}n waves propagating along the external magnetic field, it
is found that linearly the predictions agree with the one-fluid
spin model. Nonlinearly, however, the induced density fluctuations
of the spin-ponderomotive force is significantly different for the
two-spin populations. As a consequence, the self-nonlinearity of
the Alfv\'{e}n waves gets a large contribution from the spin
effects, even for a high-temperature plasma. In general, the
conclusion is that spin effects cannot be neglected even in
moderate-density high-temperature plasmas that normally are
regarded as perfectly classical.

Neglecting spin-spin interactions, the equations of motion are
\cite {marklund-brodin,MHD-spin,EP-spin,Ferro-spin,marklund-etal}
\begin{equation}
\partial _{t}n_{s}+\mathbf{\nabla }\cdot (n_{s}\mathbf{v}_{s})=0,
\label{eq:density}
\end{equation}
where $n_{s}$ and $\mathbf{v}_{s}$ are the density and velocity of species $s
$, $s=i,+,-$ enumerates the plasma particle species and $\pm $ denotes the
two types of electrons,
\begin{eqnarray}
&&\!\!\!\!m_{s}n_{s}\left( \partial _{t}+\mathbf{v}_{s}\cdot \mathbf{\nabla }%
\right) \mathbf{v}_{s}=q_{s}n_{s}\left( \mathbf{E}+\mathbf{v}_{s}\times
\mathbf{B}\right) -c_{s}^{2}\mathbf{\nabla }n_{s}  \notag \\
&&\quad +\frac{2\mu _{s}n_{s}}{\hbar }S_{s}^{a}\mathbf{\nabla }B_{a}+\frac{%
\hbar ^{2}n_{s}}{2m_{s}}\mathbf{\nabla }\left( \frac{\nabla ^{2}n_{s}^{1/2}}{%
n_{s}^{1/2}}\right) ,  \label{eq:mom-q}
\end{eqnarray}
and
\begin{equation}
\left( \partial _{t}+\mathbf{v}_{s}\cdot \mathbf{\nabla }\right) \mathbf{S}%
_{s}=-\frac{2\mu _{s}}{\hbar }\mathbf{B}\times \mathbf{S}_{s}
\label{Eq:Spin-evolve}
\end{equation}
where $\mathbf{S}_{s}$ is the spin of species $s$, $q_{s}$ is the charge of
species $s$, $p_{s}=p_{s}(T_{s};n_{s})$ is the pressure of species $s$ and $%
c_{s}=(dp_{s}/dn_{s})^{1/2}$ is the sound speed of species $s$
(where we have assumed an isothermal plasma) containing also
contributions due to the Fermi pressure, $\mu _{s}$ is the
magnetic moment of species $s$, and $-\mu _{\pm }\equiv \mu
_{B}=e\hbar /2m_{e}$ is the Bohr magneton, $e$ is the magnitude of
the electron charge, $\hbar $ is Planck's constant, $m_{e}$ is the
electron rest mass, and $c$ is the speed of light. We note that
Einstein's summation convention has been used in (2). In what follows we will, due to the large ion inertia, neglect the quantum
corrections to the ion motion.

For temporal variations of the magnetic field faster than the
inverse electron cyclotron frequency, spin flips can be induced.
Furthermore, particle collisions can also reverse the spin. Thus,
to make sure that spin reversal does not occur, we consider
dynamics on a time scale shorter than the inverse collision
frequency, but longer than the inverse cyclotron frequency. For
this case, we can replace the spin evolution equation
(\ref{Eq:Spin-evolve}) with the relation $\mathbf{S}_{\pm }=\mp
(\hbar /2)\hat{\mathbf{B}}$, for electrons with spin up and down
relative to the external magnetic field.

The coupling between the quantum plasma species is mediated by the
electromagnetic field. The magnetizations due to the different
spin sources are $\mathbf{M}_{\pm }=-2\mu _{B}n_{\pm
}\mathbf{S}_{\pm }/\hbar =\pm \mu _{B}n_{\pm }\hat{\mathbf{B}}$.
Amp\`{e}re's law then takes the form
\begin{equation}
c^{2}\mathbf{\nabla}\times \mathbf{B}=c^{2}\mu _{0}\Big[\mathbf{j}+\mathbf{%
\nabla}\times (\mathbf{M}_{+}+\mathbf{M}_{-})\Big]+\partial _{t}\mathbf{E},
\label{Eq-ampere}
\end{equation}
where the free current is denoted $\mathbf{j}$. Moreover
\begin{equation}
\varepsilon _{0}\mathbf{\nabla}\cdot \mathbf{E}=q_{i}n_{i}-e(n_{+}+n_{-}).
\end{equation}
The system is closed by Faraday's law
\begin{equation}
\mathbf{\nabla}\times \mathbf{E}=-\partial _{t}\mathbf{B}.
\label{Eq-Faraday}
\end{equation}


In previous works \cite
{marklund-brodin,MHD-spin,EP-spin,Ferro-spin,marklund-etal}, electrons have
been treated as a single population, with a single macroscopic velocity $%
\mathbf{v}$ and spin vector $\mathbf{S}$. As argued above, for fast dynamics
in an approximately collisionless plasma, this is not appropriate, as the
populations with positive and negative spins along the magnetic field will
not change spins on the short time scales considered, and as seen in Eq. (%
\ref{eq:mom-q}) the two populations are described by separate evolution
equations. If we describe the plasmas by a single electron population, with
a background spin distribution close to thermodynamic equilibrium, the spin
effects are limited to a certain extent whenever $\mu
_{B}B_{0}/k_{B}T_{e}\ll 1$. This is due to the thermodynamic Brillouin
distribution for spins $\propto \tanh (\mu _{B}B_{0}/k_{B}T_{e})$ describing
the macroscopic net effect of the spin orientation. Thus within the single
electron fluid model, we need low temperatures or very strong magnetic
fields for spin effects to be important. By contrast, within the two-fluid
electron model, spin effects may be of importance also in a weakly
magnetized high-temperature plasma, as will be demonstrated below.

As an example, we consider the nonlinear response to a
low-frequency electromagnetic Alfv\'{e}n wave pulse propagating
parallel to an external magnetic field. In linear ideal
magnetohydrodynamic (MHD) theory, the magnetic field perturbation
thus propagates along the external magnetic field
$\mathbf{B}_{0}=B_{0}\widehat{\mathbf{z}}$ with the Alfv\'{e}n
velocity $c_{A}=(B_{0}^{2}/\mu _{0}\rho _{0})^{1/2}$, where $\rho
_{0}$ is the unperturbed mass density. Since linearly the
Alfv\'{e}n wave has no density perturbation, the quantum effects
of the Bohm--de Broglie potential and the Fermi pressure do not
change this result. In what follows, such quantum effects will be
omitted. As found in \cite{Ferro-spin}, within a
single fluid spin model, the Alfv\'{e}n velocity is decreased by a factor $%
1+(\hbar \omega _{pe}^{2}/2m_{i}c^{2}\omega _{ce}^{(0)})\tanh (\mu
_{B}B_{0}/k_{B}T_{e})$ due to the spin where $\omega _{pe}$ is the
electron plasma frequency, $m_{i}$ the ion mass, $\omega
_{ce}^{(0)}=e\mu_0H_0/m_{e}$ is the electron cyclotron frequency
due to the magnetic field $\mu_0H_0 \equiv B_{0}-\mu _{0}M_{0}$,
which is the field with external sources only, i.e. the
contribution from the spins are excluded (here $M_0$ is the
unperturbed magnetization due to the spin sources). For $\mu
_{B}B_{0}\ll k_{B}T_{e}$ the correction factor for the Alfv\'{e}n
velocity is close to unity, and the approximation $\mu_0H_0
\approx B_{0}$ is a good one. This is the case that will be
considered below, and the spin corrections to the linear
Alfv\'{e}n velocity will therefore be omitted in what follows.
Furtermore, the envelope of a weakly modulated
Alfv\'{e}n wave
will propagate with a group velocity $v_{g}\simeq c_{A}$ for frequencies $%
\omega \ll \omega _{ci}$ \cite{Cyclotron-ref}, where $\omega
_{ci}$ is the ion cyclotron frequency.

The ponderomotive force of this envelope will drive low-frequency
longitudinal perturbations (denoted by superscript `lf' in what
follows) that are second order in an amplitude expansion, and to
leading order depend on a single coordinate $\zeta =z-c_{A}t$. \
Thus the dynamics is considered to be slow in a system comoving
with the group velocity. Neglecting spin, for a slowly varying
magnetic field perturbation of the
form $\mathbf{B}=B(z,t)\exp [i(kz-\omega t)]\widehat{\mathbf{e}}$ $+\mathrm{%
c.c.}$(where $\widehat{\mathbf{e}}$ is a unit vector perpendicular to $%
\widehat{\mathbf{z}}$ and $\mathrm{c.c.}$ denotes complex conjugate), the
low-frequency MHD momentum balance can be written
\begin{equation}
\rho _{0}\partial _{t}v_{i}^{\mathrm{lf}}=-\partial _{z}\left[ \left| B\right|
^{2}/\mu _{0}+(k_{B}T_{i}+k_{B}T_{e})n^{\mathrm{lf}}\right] ,\label{Eq:MHD-ponderomotive}
\end{equation}
where index $i$ denotes ions. For simplicity we assume a weak magnetic
field, $B_{0}\ll \lbrack \mu _{0}\rho
_{0}(k_{B}T_{i}+k_{B}T_{e})/m_{i}]^{1/2}$. Relating the low-frequency
perturbations of the density and velocity using Eq. (\ref{eq:density}), the
left hand side of Eq.\ (\ref{Eq:MHD-ponderomotive}) is then found to be
negligible, and the density depletion is given by
\begin{equation}
n^{\mathrm{lf}}=-\frac{\left| B\right| ^{2}}{(k_{B}T_{i}+k_{B}T_{e})\mu _{0}}
\label{Eq:density-ideal}
\end{equation}
Moreover, since we have charge neutrality within the MHD approximation, we
have here neglected the index $i$ on the density, since the total electron
density will be the same for a proton-electron plasma. Next, we add the spin
terms in our model. Again neglecting the ion inertia, the MHD low-frequency
force balance Eq. (\ref{Eq:MHD-ponderomotive}) is replaced by
\begin{equation}
F_{\mathrm{p+}}+F_{\mathrm{p-}}=\partial _{z}\left[ |B|^{2}/\mu
_{0}+(k_{B}T_{i}+k_{B}T_{e})n^{\mathrm{lf}}\right]   \label{eq:lf-equation}
\end{equation}
where the ponderomotive force contributions $F_{\mathrm{p\pm }}$ are
low-frequency perturbations due to the terms $2\mu _{s}n_{s}S_{s}^{a}\mathbf{%
\nabla }B_{a}/\hbar $ in the electron momentum equations. Including the spin
vectors component in the direction of the perturbed magnetic field, the sum
of these spin force contributions can be written
\begin{equation}
F_{\mathrm{p+}}+F_{\mathrm{p-}}\approx \frac{n_{0}}{2}\frac{\mu _{B}B_{0}}{%
k_{B}T_{e}}\frac{e\hbar }{m_{e}}\frac{\partial }{\partial z}\left( \frac{%
\left| B\right| ^{2}}{B_{0}}\right)   \label{eq:sum-pond1}
\end{equation}
where we have used that the unperturbed density difference $(n_{0+}-n_{0-})$
of the two spin populations in thermodynamical equilibrium is proportional
to $\tanh (\mu _{B}B_{0}/k_{B}T_{e})\approx \mu _{B}B_{0}/k_{B}T_{e}$. Thus
the net effect of the spin ponderomotive force on the ion density as well as
the \textit{total} electron and ion density is very small in the regime $\mu
_{B}B_{0}/k_{B}T_{e}\ll 1$, similarly as we would have for a single electron
spin model, and consequently Eq. (\ref{Eq:density-ideal}) is a valid approximation.

The interesting difference between different fluid models comes
when we analyze the density perturbations of the two electron
populations separately. The low-frequency momentum balance
equation for the different electron species are
\begin{equation}
en_{0\pm }\frac{\partial \Phi ^{\mathrm{lf}}}{\partial z}-k_{B}T_{e}\frac{%
\partial n_{\pm }^{\mathrm{lf}}}{\partial z}-\left( 1\mp \frac{e\hbar \mu
_{0}n_{0\pm }}{m_{e}B_{0}}\right) \frac{\partial }{\partial z}\left( \frac{%
\left| B\right| ^{2}}{\mu _{0}}\right) =0  \label{eq:pond-plus-minus}
\end{equation}
where we have introduced the electrostatic low-frequency potential $\Phi ^{%
\mathrm{lf}}$ (this potential does not appear in the overall
momentum balance, as the plasma is quasi-neutral). By adding the
$\pm $ parts of Eq. (\ref {eq:pond-plus-minus}) and integrating we
obtain
\begin{equation}
n_{+}^{\mathrm{lf}}+n_{-}^{\mathrm{lf}}=\frac{en_{0}\Phi ^{\mathrm{lf}}%
}{k_{B}T_{e}}-\frac{1}{k_{B}T_{e}}\left( 1-\frac{\mu _{B}B_{0}}{%
k_{B}T_{e}}\frac{\mu _{B}B_{0}}{m_{i}c_{A}^{2}}\right) \frac{\left| B\right|
^{2}}{\mu _{0}}  \label{eq:sum-density}
\end{equation}
to first order in the expansion parameter $\mu
_{B}B_{0}/k_{B}T_{e}$. Due to the small factor $\mu
_{B}B_{0}/k_{B}T_{e}$ in front of the last term, the spin effects
on the total electron population are small, in agreement with Eq.
(\ref {eq:sum-pond1}), again justifying the omission of spin
effects in Eq. (\ref {Eq:density-ideal}). However, solving instead
for the density difference between the two electron populations we
find
\begin{equation}
n_{+}^{\mathrm{lf}}-n_{-}^{\mathrm{lf}}=\frac{2}{k_{B}T_{e}}\frac{\mu
_{B}B_{0}}{m_{i}c_{A}^{2}}\frac{\left| B\right| ^{2}}{\mu _{0}}
\label{eq:density-diff}
\end{equation}

The importance of the density difference displayed in (\ref{eq:density-diff}) appears
when the nonlinear self-interaction of the Alfv\'{e}n waves is
studied. The momentum equation
contains the term $(\mathbf{j}\times \mathbf{B}/n)^{\mathrm{nl}}$, where
$\mathrm{nl}$ denotes the nonlinear part and $\mathbf{j}$ is
determined from Eq. (\ref {Eq-ampere}) with the displacement current
neglected. Omitting the terms of higher order in the
expansion parameter $\mu _{B}B_{0}/k_{B}T_{e}$ we find
\begin{equation}
\left[ \frac{\mathbf{j}\times \mathbf{B}}{n}\right] ^{%
\mathrm{nl}}\!\!\!=-\frac{(i\mathbf{k}\times \mathbf{B}_{1})\times \mathbf{B}%
_{0}}{\mu _{0}n_{0}}\frac{n^{\mathrm{lf}}}{n_{0}}\left[ 1-\left( \frac{2\mu
_{B}B_{0}}{m_{i}c_{A}^{2}}\right) ^{2}\right]   \label{eq:final}
\end{equation}
where the last term represent the two-fluid electron spin contribution. This
results in a corresponding spin-modification of the self-nonlinearity of the
Alfv\'{e}n waves. Including weakly dispersive effects due to the Hall
current, parallel propagating Alfv\'{e}n waves are described by the so
called derivative nonlinear Schr\"{o}dinger (DNLS) equation \cite
{Mjolhus1988}. For a quasi-monochromatic wave as considered here, the DNLS
equation reduces to the usual nonlinear Schr\"{o}dinger (NLS) equation \cite
{Brodin1996} of the form
\begin{equation}
i\partial _{t}B_{1}+\frac{v_{g}^{\prime }}{2}\partial _{\zeta }^{2}B_{1}+Q%
\frac{\left| B_{1}\right| ^{2}}{B_{0}^{2}}B_{1}=0  \label{eq:nls}
\end{equation}
Here the group dispersion $v_{g}^{\prime }=dv_{g}/dk$ is the group
dispersion, $\zeta =z-v_{g}t$ is the comoving coordinate and $v_{g}$ is the
group velocity. These quantities are determined from the Alfv\'{e}n wave
dispersion relation, which reads $\omega ^{2}=k^{2}c_{A}^{2}(1\pm
kc_{A}/\omega _{ci})$, when weakly dispersive effects due to the Hall
current is included \cite{Cyclotron-ref}. The upper (lower) sign corresponds
to right (left) hand circular polarization. The nonlinear coefficient is $%
Q=Q_{c}[1-({2}${$\mu _{B}B_{0}$}$/{m_{i}c_{A}^{2}})^{2}]$, where
the classical coefficient is
$Q_{c}=kc_{A}^{3}/4(c_{A}^{2}-c_{s}^{2})$ $\simeq
-kc_{A}^{3}/4c_{s}^{2}$. The NLS equation has been studied
extensively \cite {Dodd-wave}, and as is well-known it admit
soliton solutions in 1D, and can describe nonlinear self-focusing
followed by collapse in higher dimensions. Furthermore, the
evolution depends crucially on the sign of the nonlinear
coefficent, which may change due to the spin effects. However, our
main concern in this context is not the evolution of the
Alfv\'{e}n waves, which were chosen just as an illustration. The
fact that interests us here is that spin can modify the dynamics
even when the spins are almost randomly distributed due to a
moderately high temperature (i.e. when $\mu
_{B}B_{0}/k_{B}T_{e}\ll 1$). The approximately random distribution
of spins is shown in the dispersion relation of the linear wave
modes, which are more or less unaffected by the spins since
linearly the total spin contribution on the electrons cancel to
leading order. Nonlinearly, however, the consequences of the
different density fluctuations induced in the spin-up and
spin-down populations are seen. The unique feature of this quantum
effect is that is survives even for a high temperature. By
contrast, well-known quantum plasma effects like the Fermi
pressure, and the Bohm--de Broglie potential becomes insignificant
for high temperatures. This is also true for single fluid spin
effects \cite
{marklund-brodin,MHD-spin,EP-spin,Ferro-spin,marklund-etal}.

An illustration of the regimes where the different quantum plasma
effects become significant is provided in Fig. 1. In particular we
note that the two-fluid nonlinear spin effects are important for
high plasma densities and/or a weak (external) magnetic fields.
For comparison, both the Fermi pressure and the Bohm--de Broglie
potential need a low-temperature or a very high density to be
significant. Single-fluid spin effects can also be significant in
this regime, or in the regime of ultra-strong magnetic fields that
can occur in astrophysical applications. Especially interesting is
that two-fluid nonlinear spin effects can be significant in a
high-temperature regime that is normally perceived as classic.
While this obviously is an intriguing result, a word of caution is
needed. Although our results clearly show that spin effects can be
important when $\mu _{B}B_{0}/m_{i}c_{A}^{2}$ approaches unity, we
note that in a number of applications, spin effects of the kind
discussed here can be suppressed even if $\mu
_{B}B_{0}/m_{i}c_{A}^{2}$ is large. These include:

--- Systems where the dynamics is dominated by compressional effects. In
such problems the thermal pressure force dominates over the spin force of
the electrons, and spin effects are suppressed unless $\mu
_{B}B_{0}/k_{B}T_{e}\rightarrow 1$;

--- High density collisional plasmas. In such cases the spin orientation
changes rapidly and different spin populations cannot be sustained;

--- Systems where the spin forces are negligible as compared to the
electrostatic force.

\begin{figure}[tbp]
\centering\includegraphics[width=\columnwidth]{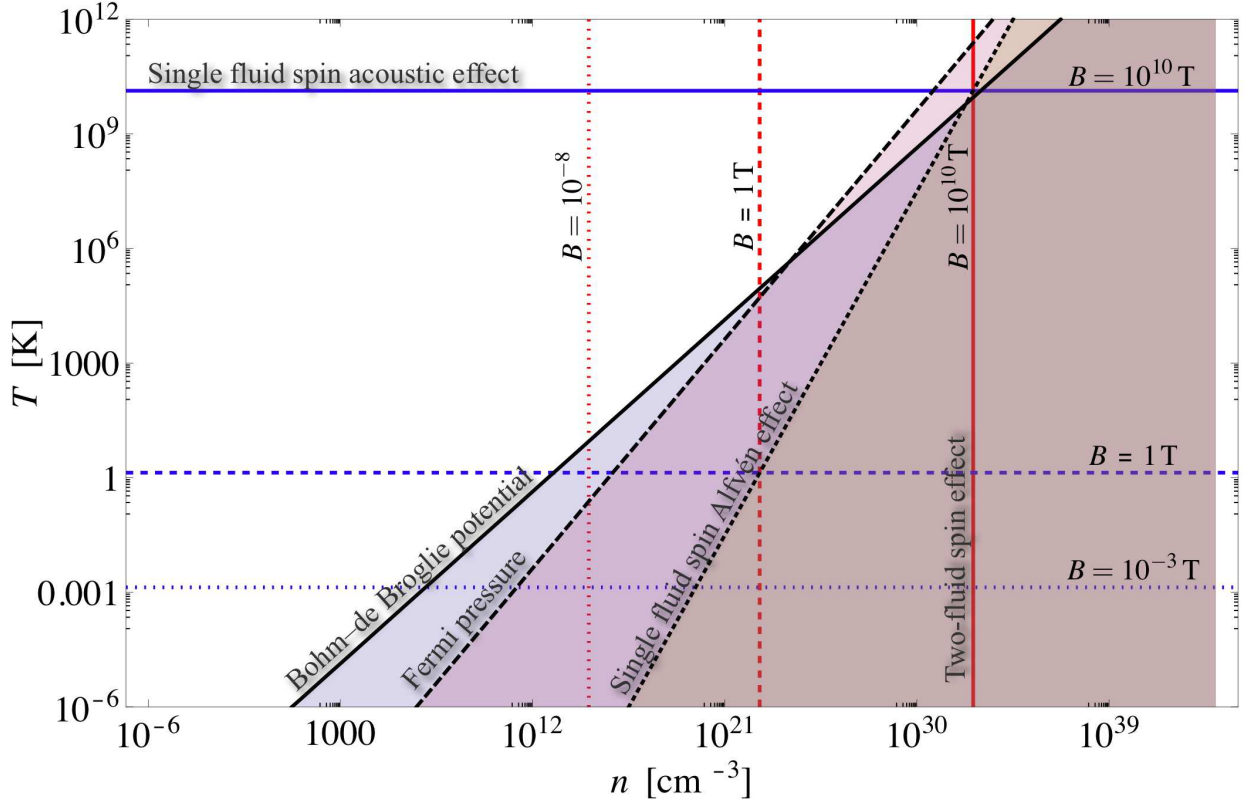}
\caption{Regions of importance in parameter
space for various quantum plasma effects. The lines are defined by
different dimensionless quantum parameters being equal to unity.
The effects included in the figure are the Fermi pressure (dashed black curve), described by the parameter $T_F/T_e \propto
\hbar^{2}n_{0}^{2/3}/m k_{B}T_e$, the Bohm--de
Broglie potential (solid black curve), described by the parameter $\hbar
\protect\omega _{pe}/k_{B}T_e \propto n^{1/2}/T_e$, 
single-fluid spin Alfv\'{e}n effects (dotted black curve), described by the parameter
$\hbar ^{2}\protect\omega _{pe}^{2}/mc^{2}k_{B}T_e$, and 
single-fluid spin acoustic effects (horizontal blue curves), described by the parameter
$\mu_{B}B_{0}/k_{B}T_e$, where three different magnetic
field strengths are depicted. The quantum regime correspond to lower
temperatures, i.e. it exists below each of the three horizontal
curves. Lastly, the two-fluid spin nonlinear effect derived in this Letter, described by the
parameter $\mu_{B}B_{0}/m_{i}c_{A}^{2} \propto n/B_0$, is depicted by the three vertical
red curves. The quantum
regime corresponds to higher densities, i.e. it exists to the
right of each of the three vertical lines.}
\end{figure}

Furthermore, the strong magnetic fields needed for confinement
tend to make the two-fluid spin effects studied here negligible
for magnetically confined plasmas. Nevertheless, even with the
above cases excluded, our discussion shows that there is a large
range of plasma problems that traditionally have been dealt with
using purely classical plasma equations, but where the electron
spin properties give a significant contribution to the dynamics.
In general the mechanism can be summarized as follows: For a
weakly magnetized initially homogeneous plasma, the spin-up and
down populations are (approximately) equal. However, when an
electromagnetic perturbation enters the system, the spin
ponderomotive force separates the two populations, which in turn modifies
the magnetic field since spin-magnetization no longer cancels. From then on, a
two-fluid electron model is needed. In particular, in the region
of aligned electron spins, the original magnetic field will be
enhanced, and hence mechanisms of this type can play the role of a
magnetic dynamo. Suitable plasma conditions for nonlinear
two-fluid spin effects to be important may be found in weakly
magnetized inertially confined plasmas, near atmospheric pressure
plasma dischages, weakly magnetized regions of the sun's
convection zone, as well as in astrophysical and cosmological
plasmas. Exploration of these vast range of problems remain a
major research project.

\acknowledgments
MM and GB was supported by the Swedish Research Council.



\begin{thebibliography}{99}
\bibitem{manfredi}  G.\ Manfredi, Fields Inst. Comm. \textbf{46}, 263 (2005).
Also available at arXiv:quant-ph/0505004.

\bibitem{haas-etal1}  F. Haas, G. Manfredi, and M. R. Feix, Phys. Rev. E
\textbf{62}, 2763 (2000).

\bibitem{haas}  F. Haas, Phys. Plasmas \textbf{12}, 062117 (2005).

\bibitem{haas-etal2}  F. Haas, L. G. Garcia, J. Goedert, and G. Manfredi,
Phys. Plasmas \textbf{10}, 3858 (2003).

\bibitem{garcia-etal}  L. G. Garcia, F. Haas, L. P. L. de Oliviera, and J.
Goedert, Phys. Plasmas \textbf{12}, 012302 (2005).

\bibitem{shukla}  P. K. Shukla, Phys. Lett. A \textbf{352}, 242 (2006).

\bibitem{Shukla-Eliasson}  P. K. Shukla and B. Eliasson, Phys. Rev. Lett.
\textbf{96}, 245001 (2006).

\bibitem{marklund-brodin}  M.\ Marklund and G.\ Brodin, Phys.\ Rev.\ Lett.\
\textbf{98}, 025001 (2007).

\bibitem{MHD-spin}  G. Brodin and M. Marklund, New J. Phys. \textbf{9}, 277
(2007)

\bibitem{EP-spin}  G. Brodin and M. Marklund, Phys. Plasmas \textbf{14},
112107 (2007)

\bibitem{Ferro-spin}  G. Brodin and M. Marklund, Phys. Rev. E \textbf{76},
055403(R) (2007)

\bibitem{marklund-etal}  M. Marklund, B. Eliasson, and P. K. Shukla, Phys.
Rev. E \textbf{76}, 067401 (2007).

\bibitem{brodin-etal}  G. Brodin, M. Marklund, B. Eliasson, and P. K.
Shukla, Phys. Rev. Lett. \textbf{98}, 125001 (2007).

\bibitem{piazza}  A. Di Piazza, K. Z. Hatsagortsyan, C. H. Keitel, Phys.
Plasmas, \textbf{14}, 032102 (2007).

\bibitem{Lundin}  J. Lundin, J. Zamanian, M. Marklund, and G. Brodin, Phys.
Plasmas, \textbf{14}, 062112 (2007).

\bibitem{Cowley}  S. C. Cowley, R. M. Kulsrud, and E. Valeo, Phys. Fluids
\textbf{29}, 430 (1986)

\bibitem{Kulsrud}  R. M. Kulsrud, E. J. Valeo, and S. C. Cowley, Nucl.
Fusion \textbf{26}, 1443 (1986).

\bibitem{Ferrofluid}  R. E. Rosensweig, Ferrohydrodynamics (Cambridge
University Press, Cambridge, 1985).

\bibitem{RMP}  M. Marklund and P. K. Shukla, Rev. Mod. Phys. \textbf{78, }591%
\textbf{\ }(2006).

\bibitem{SPP}  M. Marklund, G. Brodin, L. Stenflo, and C. S. Liu, Phys. Rev.
Lett., submitted (2007) (arXiv:0712.3145).

\bibitem{Quantum wells}  G. Manfredi and P.-A. Hervieux, Appl. Phys. Lett.
\textbf{91}, 061108 (2007).

\bibitem{Ultracold}  W. Li, P. J. Tanner, and T. F. Gallagher, Phys. Rev.
Lett. \textbf{94}, 173001 (2005).

\bibitem{Harding2006}  A. K. Harding and D. Lai, Rep. Prog. Phys. \textbf{69}%
, 2631 (2006).

\bibitem{Glenzer2007}  S. H. Glenzer \textit{et al.}, Phys. Rev. Lett.
\textbf{98}, 065002 (2007).

\bibitem{Cyclotron-ref}  G. Brodin and L. Stenflo, Contrib. Plasma Phys.
\textbf{30}, 413 (1990).

\bibitem{Mjolhus1988}  E. Mj\"{o}lhus and J. Wyller, J. Plasma Phys.,
\textbf{40}, 299, (1988).

\bibitem{Brodin1996}  G. Brodin, J. Plasma Phys., \textbf{55}, 121, (1996).

\bibitem{Dodd-wave}  R. K. Dodd, J. C. Eilbeck, J. D. Gibbon and H. C.
Morris, \textit{Solitons and Nonlinear Wave Equations} (Academic Press,
London, 1982).





































\end{thebibliography}
\end{document}